\documentclass[12pt,preprint]{aastex}
\usepackage{graphicx,multirow}
\usepackage{natbib}

\long\def\symbolfootnotetext[#1]#2{\begingroup%
\def\thefootnote{\fnsymbol{footnote}}\footnotetext[#1]{#2}\endgroup} 

\shorttitle{}
\shortauthors{\"Oberg et al.}

\begin{document}

\title{Complex molecules toward low-mass protostars: the Serpens core$^*$\symbolfootnotetext[1]{Based on observations carried out with the IRAM 30-m telescope. IRAM is supported by INSU/CNRS (France), MPG (Germany) and IGN (Spain).}
}

\author{Karin I. \"Oberg\altaffilmark{1}}
\affil{Harvard-Smithsonian Center for Astrophysics, MS 42, 60 Garden St, Cambridge, MA 02138, USA.}

\author{Nienke van der Marel, Lars E. Kristensen, Ewine F. van Dishoeck\altaffilmark{2}}
\affil{Leiden Observatory, Leiden University, P.O. Box 9513, 2300 RA Leiden, The Netherlands}

\altaffiltext{1}{Hubble fellow.}
\altaffiltext{2}{Max-Planck Institute f{\"u}r Extraterrestrische Physik, Giessenbachstr. 1, 85748 Garching, Germany}

\begin{abstract}
\noindent Gas-phase complex organic molecules are commonly detected toward high-mass protostellar hot cores. Detections toward low-mass protostars and outflows are comparatively rare, and a larger sample is key to investigate how the chemistry responds to its environment. Guided by the prediction that complex organic molecules form in CH$_3$OH-rich ices and thermally or non-thermally evaporate with CH$_3$OH, we have identified three sight-lines in the Serpens core -- SMM1, SMM4 and SMM4-W -- which are likely to be rich in complex organics. Using the IRAM 30m telescope, narrow lines (FWHM of 1--2 km s$^{-1}$) of CH$_3$CHO and CH$_3$OCH$_3$ are detected toward all sources, HCOOCH$_3$ toward SMM1 and SMM4-W, and C$_2$H$_5$OH not at all. Beam-averaged abundances of individual complex organics range between 0.6 and 10\% with respect to CH$_3$OH when the CH$_3$OH rotational temperature is applied.  The summed complex organic abundances also vary by an order of magnitude, with the richest chemistry toward the most luminous protostar SMM1. The range of abundances compare well with other beam-averaged observations of low-mass sources. Complex organic abundances are of the same order of magnitude toward low-mass protostars and high-mass hot cores, but HCOOCH$_3$ is relatively more important toward low-mass protostars. This is consistent with a sequential ice photochemistry, dominated by CHO-containing products at low temperatures and early times.

\end{abstract}


\section{Introduction}

Complex organic molecules in space are traditionally associated with hot cores in high-mass protostars. IRAS 16293-2422 seems to be a low-mass equivalent to the high-mass hot cores \citep{vanDishoeck95,Cazaux03}, and there are at least three other low-mass protostars
where complex molecules such as HCOOCH$_3$ and CH$_3$OCH$_3$ are detected \citep[e.g.][]{Bottinelli04b, Jorgensen05}. A low-mass outflow has also been detected in a number of  organic species \citep{Arce08} and HCOOCH$_3$ and CH$_3$CHO were recently detected in  a cold and quiescent part of a low-mass star forming region that probably has been illuminated by UV through an outflow cavity \citep{Oberg10a}. Complex organic molecules are clearly present in a range of environments, illustrating the presence of efficient pre-biotic pathways to chemical complexity \citep[see][for a review]{Herbst09}.

To explain the observed abundances of complex organics (1-50\% with respect to CH$_3$OH), and their formation in environments with orders of magnitude differences in density, temperature, radiation flux and age, most models invoke grain surface chemistry. \citet{Garrod06} and \citet{Garrod08} set up a model framework, where complex molecules form in ices from recombination of photodissociation fragments of simpler species, especially CH$_3$OH ice. The fragments become mobile in the protostellar phase as the grains are falling in toward the protostars, forming larger and larger molecules that eventually evaporate. This scheme can reproduce many aspects of the molecular abundance patterns observed in hot core regions. An ice formation scenario is also consistent with a hot core survey \citep{Bisschop07}, where most complex molecules are closely correlated with the grain surface product CH$_3$OH. 

Laboratory experiments confirm that UV-induced CH$_3$OH ice chemistry is a feasible way of producing large amount of the observed hot-core molecules \citep[e.g.][]{Gerakines95,Oberg09d}. The experiments also reveal that non-thermal diffusion of radicals may be possible in the ice before the onset of thermal diffusion, resulting in a cold $T<20$~K production channel of complex organics. Since UV photodesorption is efficient \citep{Oberg09b,Oberg09c}, many gas-phase complex molecules toward protostars may be present in the envelope or in outflow cavities, and not only in the innermost hot core where ices are thermally evaporated.

Low-mass star forming regions are less complicated than the high-mass equivalents and are therefore a good test-bed for theories on complex molecule formation. Because of the small sample of low-mass sources (six total) with complex molecule detections, it has so far been difficult to ascertain 1) whether there is a real difference between the chemical evolution in low- and high-mass protostars and 2) if the predictions coming out of laboratory experiments and models are supported by observations. The Serpens molecular core is a prime target to search for complex molecules and thus expand the current sample of low-mass complex molecule sources. Several of the protostars in the core are embedded in material where CH$_3$OH makes up  30\% of the ice with respect to H$_2$O, the most CH$_3$OH-ice-rich low-mass region observed to date \citep{Pontoppidan04}. From CH$_3$OH gas observations \citep{Kristensen10}, the region experiences significant ice desorption, which is a requirement to actually observe the products of the complex grain surface chemistry. Similar conditions toward B1-b were previously used to justify a successful search for complex molecules there \citep{Oberg10a}.

Within the Serpens core we targeted the lines of sight toward the protostars SMM1 and SMM4 and toward SMM4-W, an outflow associated with SMM4. SMM1 and SMM4 have bolometric luminosities of $\sim$30 and 5 L$_\odot$, respectively assuming a distance of 250 pc \citep{vanDishoeck10}. These three lines-of-sight thus sample three distinctly different environments within the same core, with potentially different chemical evolutionary tracks. This offers a prime opportunity to address the response of the complex organic chemistry to the environment and thus provide insight to their dominating formation mechanisms. 

The paper is structured as follows. \S\ref{sec:obs} presents the observational parameters for the search of complex organics toward SMM1, SMM4 and SMM4-W. The search results are used together with CH$_3$OH column densities from \citet{Kristensen10} to derive abundances with respect to CH$_3$OH in \S\ref{sec:res}. The determined abundances are discussed with respect to the different environments within the Serpens core and the distribution of complex molecules in low- and high-mass protostellar regions more generally in \S\ref{sec:disc}. Based on these comparisons we then discuss the formation mechanisms of complex organic molecules.

\section{Observations and data reduction\label{sec:obs}}

\begin{figure}
\epsscale{0.7}
\plotone{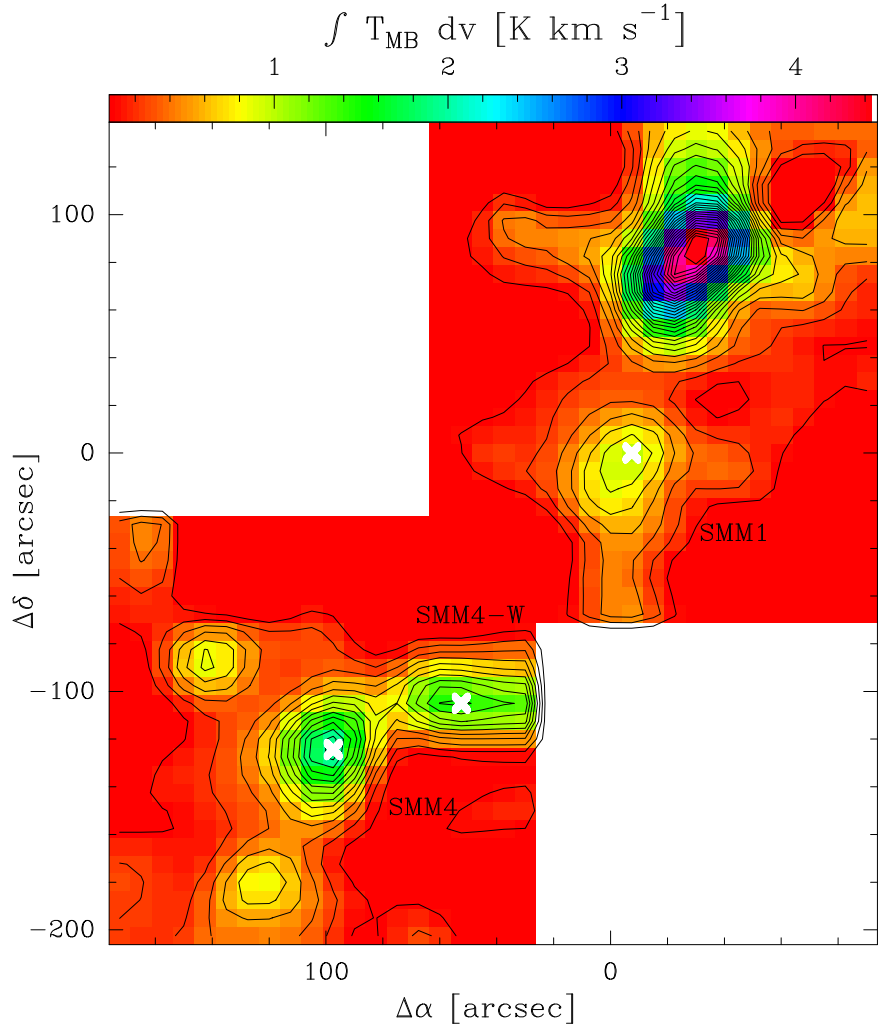}
\caption{Integrated intensity map of CH$_3$OH 7$_0$--6$_0$ A$^+$ line at 338.41 GHz. The SMM4, SMM4-W and SMM1 positions are marked with white crosses \citep{Kristensen10}.\label{fig:ch3oh}}
\end{figure}

Figure \ref{fig:ch3oh} shows a map of CH$_3$OH emission adapted from \citet{Kristensen10} with SMM1, SMM4 and SMM4-W marked. The three sources were observed with the IRAM 30-m telescope in July 2010, during good weather conditions ($\tau\sim0.2-0.3$), using the EMIR receiver. The three frequency setups in Table \ref{tab:obs} were designed to cover a maximum number of complex organic molecules, typically detected in hot cores, with the focus on lines with upper level energies of 10--100~K because of the previously discovered low excitation temperature of CH$_3$OH toward the same sources. The settings contain multiple lines for HCOOCH$_3$, CH$_3$CHO, CH$_3$OCH$_3$ and C$_2$H$_5$OH (and additional HOCH$_2$CHO and C$_2$H$_5$CN lines) to allow us to constrain the excitation temperatures of any detected species. The line frequencies are taken from the JPL molecular database and the Cologne Database for Molecular Spectroscopy \citep{Muller01}.

Each receiver setting consists of a combination of the EMIR 90 and 150 GHz receivers (Table \ref{tab:obs}). At these wavelengths, the beam sizes are $\sim$28 and 17'', respectively. All three settings were used toward SMM1 and SMM4, while only setting 1 and 2 were observed toward SMM4-W.

Each receiver was connected to a unit of the autocorrelator, with a spectral resolution of 40~kHz and a bandwidth of 120~MHz, equivalent to an unsmoothed velocity resolution of $\sim$0.1 km s$^{-1}$. All spectra were then smoothed to 0.3--0.5 km s$^{-1}$ to enhance the S/N. Typical system temperatures were 100-150 K. All observations were carried out using wobbler switching with a 100'' throw. Pointing was checked every $\sim$2 hr on Mars, J1749+096 or J1741-038 with a typical pointing accuracy of 1.5''. All intensities reported in this paper are expressed in units of main-beam brightness temperature, which were converted from antenna temperatures in {\it pako} using reported main beam and forward efficiencies (B$_{\rm eff}$ and F$_{\rm eff}$) of 75\% and 95\% with the E90 receiver and, 69\% and 93\% with E150 receiver. The rms in mK are reported in Table \ref{tab:obs} for the binned spectra. We assume a calibration uncertainty of 10\%.

The data were reduced with the CLASS program, part of the GILDAS software package (see http://www.iram.fr/IRAMFR/GILDAS). Linear (first-order) baselines were determined from velocity ranges without emission features, and then subtracted from the spectra. Some velocity channels showed spikes (VESPA spikes are a known problem, in particular in conjunction with EMIR), which were replaced by interpolating the closest two good channels. There were no spikes within 10 km s$^{-1}$ of any targeted line and replacing the bad channels is not affecting the results.

\section{Results\label{sec:res}}

\noindent This section presents detected complex molecule spectra and the inferred abundances with respect to CH$_3$OH. First, the CH$_3$OH column densities and rotational temperatures are extracted from a previously published CH$_3$OH map convolved to the E90 and E150 beam sizes. Second, complex organic lines are identified from the new observations, and integrated line intensities are derived. Third, the complex molecular column densities and abundances are calculated assuming that the level population of the complex molecules can be described by the excitation temperatures derived for CH$_3$OH.   

\subsection{CH$_3$OH columns and temperatures}

CH$_3$OH observations toward the Serpens core are presented in \citet{Kristensen10} in the form of a JCMT HARP-B map. To compare CH$_3$OH and complex molecule abundances, the fully sampled CH$_3$OH map, obtained with a 15" beam, in Fig \ref{fig:ch3oh} was convolved using MIRIAD \citep{Sault95} with the complex molecule beam sizes of $\sim$17 and 28" for the 2 and 3~mm observations respectively. The resulting CH$_3$OH line intensities at the SMM1, SMM4 and SMM4-W positions are used to derive new column densities and rotational temperatures using the rotational diagram method (Appendix A). The column densities and excitation temperatures are listed in Table \ref{tab:ch3oh}, demonstrating that the applied beam size has only a marginal impact. CH$_3$OH thus seems equally abundant on 17 and 28" scales, suggestive of that the emission is dominated by the protostellar envelope rather than a central hot core.

The derived column densities toward SMM1 and SMM4-W are up to 30\% lower compared to the results presented in \citet{Kristensen10} based on a 15" beam. This is due to under-weighting of some weak lines in  \citet{Kristensen10}, which resulted in higher rotational temperatures and higher column densities. 

\subsection{Complex molecule line detections}

\begin{figure}[htp]
\epsscale{0.7}
\plotone{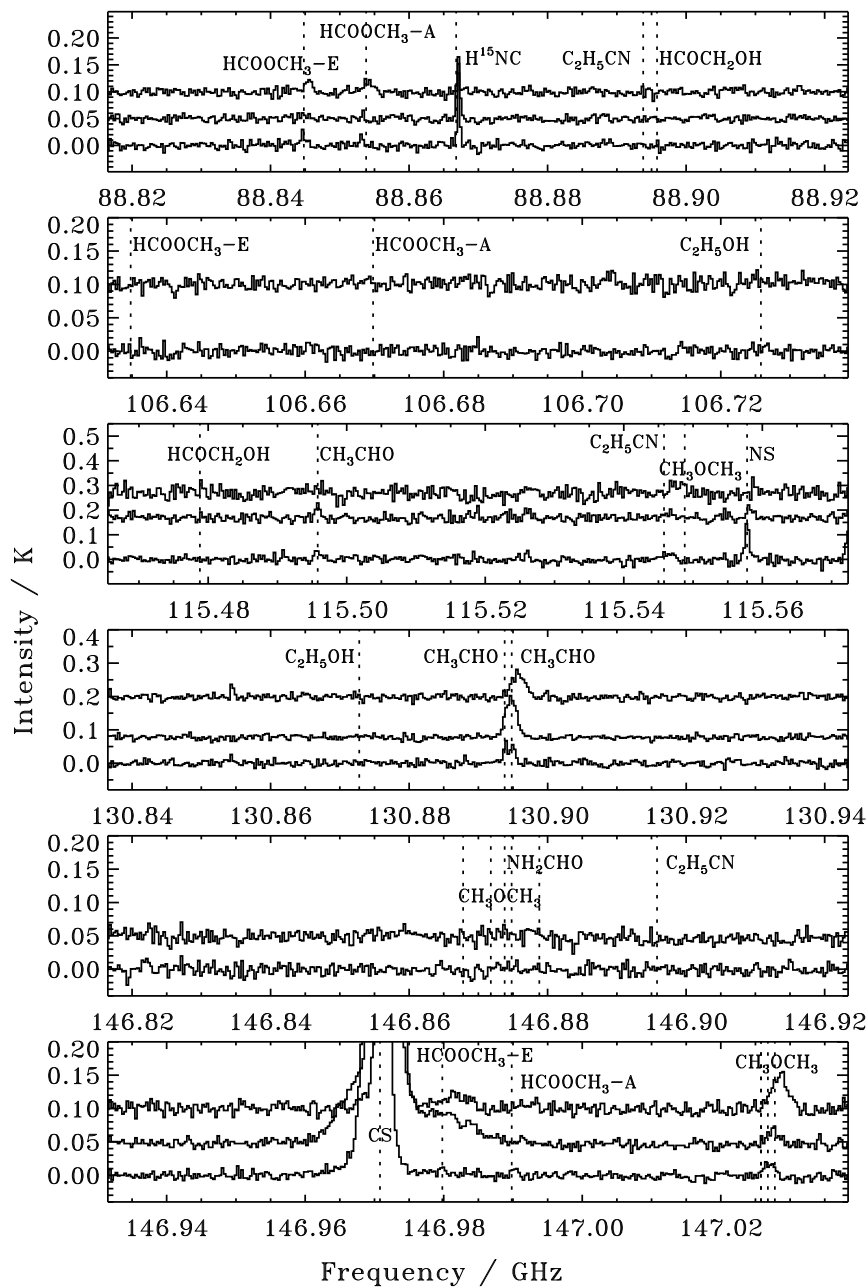}
\caption{Complete IRAM 30m spectra toward the SMM4-W outflow (upper spectra), SMM4 (middle spectra) and SMM1 (lower spectra) in the three different E90/E150 settings. The spectra toward SMM4 and SMM-W are offset for clarity. In addition to the labeled lines, there is an unidentified line at 130.855 GHz in SMM-4 W. \label{fig:comp_sp_full}}
\end{figure}

\begin{figure*}[htp]
\epsscale{1}
\plotone{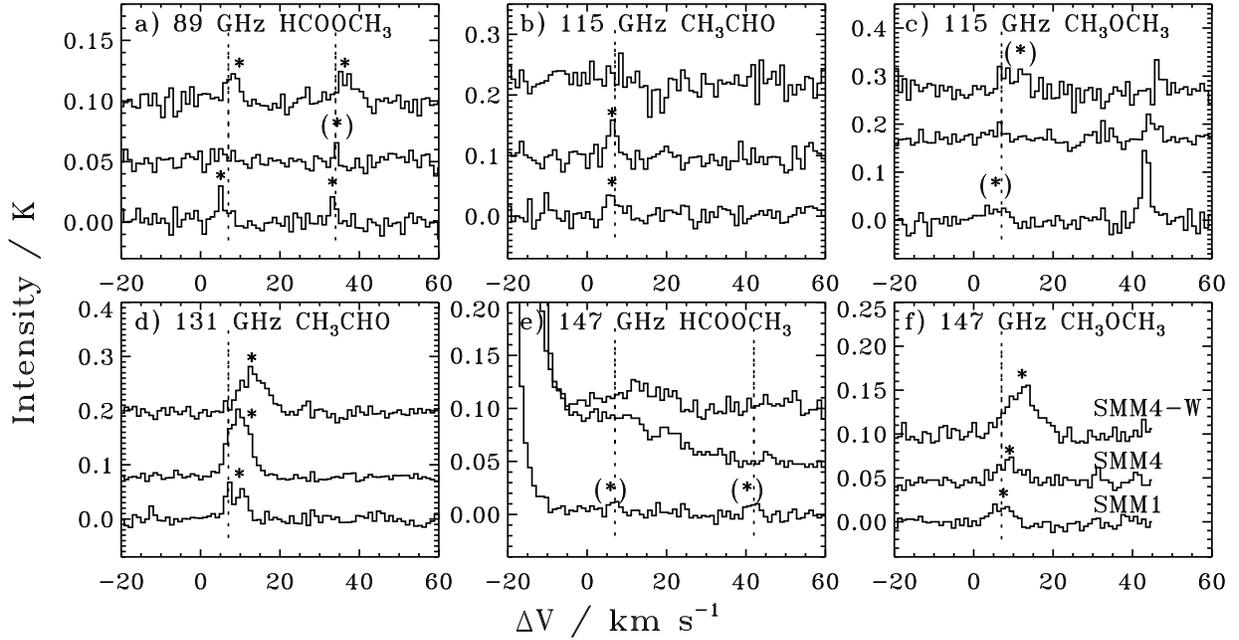}
\caption{HCOOCH$_3$, CH$_3$CHO and CH$_3$OCH$_3$ lines toward the SMM4-W outflow (upper spectra), SMM4 core (middle spectra) and SMM1 core (lower spectra). Detections are marked with stars and tentative detections with stars in parantheses. The dotted line marks the SMM1 velocity of 7 km s$^{-1}$. The spectra toward SMM4 and SMM-W are offset for clarity.\label{fig:comp_sp}}
\end{figure*}

Figure \ref{fig:comp_sp_full} shows the complete spectra for all settings. Figure \ref{fig:comp_sp} zooms in on the most diagnostic lines and shows that SMM1, SMM4 and SMM4-W all contain complex organic molecules with peak intensities of 20--100~mK. CH$_3$CHO and CH$_3$OCH$_3$ are detected in all lines of sight, while at least two HCOOCH$_3$ lines are detected toward SMM1 and SMM4-W and tentatively detected toward SMM4. C$_2$H$_5$OH, HCOCH$_2$OH and C$_2$H$_5$CN are not detected.

Table \ref{tab:comp} list all detections and significant upper limits. The FWHM vary between 0.7--2.5 toward SMM1, 1.0--2.6~km~s$^{-1}$ toward SMM4 and 1.6--4.4~km~s$^{-1}$ toward SMM4-W. Only the HCOOCH$_3$ lines around 88.8 GHz and the CH$_3$CHO line around 115 GHz are not blended, however, and these lines have FWHM of  0.7--1.2, 1.0 and 1.6--2.3~km~s$^{-1}$ toward SMM1, SMM4 and SMM4-W, respectively. These lines are  narrow compared to the CH$_3$OH FWHM of 3.4--4.0~km~s$^{-1}$ toward all three sources \citep{Kristensen10}. There is thus no evidence that the complex organics originate in a hotter or more turbulent region than CH$_3$OH, justifying our presentation of beam averaged abundances below.

The line positions of all complex molecules toward SMM1 and SMM4 are consistent with a constant source velocity, indicative of that all molecular emission has the same origin. This is not the case toward SMM4-W. The observed HCOOCH$_3$ emission has the same source velocity as SMM4, while the CH$_3$CHO and the CH$_3$OCH$_3$ emission are shifted by $\sim$5 km s$^{-1}$, comparable to the shift in CH$_3$OH emission observed by \citet{Kristensen10}. CH$_3$CHO and CH$_3$OCH$_3$ emission then probably originate in the observed CH$_3$OH outflow, while HCOOCH$_3$ emission originates in a more quiescent region associated with the extended SMM4 envelope.

The lines are integrated numerically assuming central line velocities of 7, 7 and 7--12 km s$^{-1}$, for SMM1, SMM4 and SMM4-W, respectively. 3$\sigma$ upper limits are derived assuming 1 km s$^{-1}$ line widths toward SMM1 and SMM4 and 2 km s$^{-1}$ line widths toward SMM4-W, and the standard equation  $\sigma=rms\times FWHM / \sqrt{n_{\rm ch}}$, where $n_{\rm ch}$ is the number of channels across the FWHM. The rms and resolutions are listed in Table \ref{tab:obs} and the resulting line intensities and upper limits in Table \ref{tab:comp}. The lowest upper limits are $\sim$20~mK~km s$^{-1}$, while detections range between 30 and 560~mK~km s$^{-1}$; most upper limits are significant compared to the detected line intensities. 

\subsection{Complex molecule column densities and abundances}

\noindent Assuming that the molecules are thermalized at a rotational temperature, optically thin emission and no beam dilution we can use

\begin{equation}
N = \frac{1.67\times10^{14}}{\nu\mu^2S} Q(T_{\rm rot})e^{\rm -E_u/T_{rot}}\int T_{\rm b}dv,
\end{equation}

\noindent from \citet{Thi04}, where $N$ is the total number of molecules, $Q(T_{\rm rot})$ the temperature dependent partition function, $E_{\rm u}$ the energy of the upper level in K, $T_{\rm rot}$ the rotational temperature, and $\int T_{\rm b}dv$ the integrated line intensity.  $\mu^2S$ is gathered from CDMS for each molecule or calculated from $I(300~K)$ in JPL, $Q(T_{\rm rot})$ is calculated at the CH$_3$OH rotational temperature by second order interpolation between the three closest values, $E_{\rm u}$ is calculated from the listed $E_{\rm low}$ values. Using the CH$_3$OH excitation temperature is a reasonable approximation since the thermal desorption temperatures and the non-thermal desorption efficiencies are expected to be similar for CH$_3$OH and more complex ices. Also, because of the fact that $\mu^2S$ is similar for all considered molecules, they should be excited to a similar level.

The resulting column densities are listed in Table \ref{tab:cols} and when constraints are available from multiple lines, the column densities are consistent within a factor of two. This supports the assumption that the CH$_3$OH gas and the complex molecules have a similar excitation temperature.  The abundances of complex molecules are calculated based on the derived complex molecule column densities and the CH$_3$OH column densities listed in Table \ref{tab:ch3oh}. SMM1 has the highest overall complex molecular abundances with 10\% HCOOCH$_3$, 6\% CH$_3$CHO and 5\% CH$_3$OCH$_3$, all with respect to CH$_3$OH. In the other two sources most complex abundances are $\sim$1\% with respect to CH$_3$OH.

\section{Discussion\label{sec:disc}}

\subsection{Variations in the complex molecule distributions in the Serpens core} 

The three observed positions in the Serpens core are different both with respect to their total complex molecular content, as traced by HCOOCH$_3$, CH$_3$CHO and CH$_3$OCH$_3$, and the relative abundances of the targeted molecules. The summed detected complex molecular abundance with respect to CH$_3$OH is $\sim$22\%  toward SMM1, 3\% toward SMM4 and  5\% toward SMM4-W. The abundances thus vary over almost an order of magnitude, with the most complex molecules toward the most luminous protostar. This is in line with expectations that, averaged over the envelope, more complex molecules are produced in the CH$_3$OH ices when a larger portion of the envelope is heated and exposed to more intense UV radiation.

Toward SMM1 and SMM4-W, HCOOCH$_3$ is the most abundant complex molecule, while CH$_3$CHO dominates the observed chemistry toward SMM4. In a scenario where most complex molecules originate from UV chemistry in CH$_3$OH-rich ices, both  HCOOCH$_3$ and CH$_3$CHO are expected to mainly form when cold, CO-rich ices are irradiated \citep{Oberg09d}. In contrast, CH$_3$OCH$_3$ is only expected to form efficiently from CH$_3$OH ice photochemistry once CO has evaporated. The HCOOCH$_3$/CH$_3$OCH$_3$ ratio is therefore a potential tracer of the relative importance of cold versus luke-warm complex ice chemistry. The (HCOOCH$_3$+CH$_3$CHO)/CH$_3$OCH$_3$ abundance ratios are $\sim$3 toward all three sources, providing evidence for that the cold complex chemistry dominates the beam-averaged emission -- consistent with the derived low rotational temperatures. Gas observations of ice chemistry products in such cold regions are possible due to e.g. UV photodesorption, desorption due to release of chemical energy and sputtering in shocks. 

Based on the observations, the main desorption mechanisms may differ for different complex molecules close to outflows. The low central velocities of the HCOOCH$_3$ lines toward the outflow SMM4-W suggests that it does not originate at the shock front where ices are sputtered by the shock, but rather in the surrounding colder material, where UV rays from the outflow cavity may still be able to penetrate. In contrast, the CH$_3$CHO  and CH$_3$OCH$_3$ emission coincides with the outflow (traced by e.g. CH$_3$OH). Similarly toward B1-b, HCOOCH$_3$ has no outflow signature, while CH$_3$CHO seems partially associated with a weak outflow \citep{Oberg10a}. 

This difference between HCOOCH$_3$, CH$_3$CHO and CH$_3$OCH$_3$ emission may be the result of different ice surface and ice mantle compositions together with the experimental results that UV photodesorption is only active in the ice surface layer \citep{Oberg09b,Oberg09c}, while the shock front results in desorption of the entire ice mantle.  If HCOOCH$_3$ mainly forms on ice surfaces, low levels of HCOOCH$_3$ gas should be present throughout the protostellar envelopes. Its relative importance to other complex molecules should be less in regions where the entire ice mantle has desorbed at e.g. shock fronts and inside hot cores. Determining the mechanism behind the different ice surface and ice mantle compositions requires detailed modeling beyond the scope of this paper. Below we instead aim to provide more empirical constraints on the ice chemistry evolution by comparing our main results with previous observations toward low- and high-mass protostars. 

\subsection{Comparison with previous complex chemistry observations}

Table \ref{tab:lowmass} lists the low-mass sources toward which at least one hot-core type molecule has been detected. The list comprises one illuminated cold core edge, two outflows, six protostellar envelopes, and three low-mass hot cores observed with 2" beams. Most previous single-dish studies assumed some level of beam dilution based on the assumption that the complex molecules mainly originate from thermal desorption close to the protostar. \citet{Oberg10a} showed that abundant HCOOCH$_3$ may instead be due to non-thermal desorption at large scales, and {\it a priori} it is difficult to tell whether the complex organics observed with single dish telescopes toward low-mass YSOs mainly reside in a hot core or are spread out in the envelope. We have therefore use abundances derived under the assumption that CH$_3$OH and complex organics have the same distributions \citep{Herbst09}.

The summed detected complex molecule abundance are 2.5--6\% toward the outflows (B1-b, SMM4-W, L1157), 9--56\% toward the protostellar envelopes (excluding SMM4), where emission from the cold and hot regions are blended, and 2--3\% toward the low-mass hot cores IRAS~16293A, IRAS~16293B, IRAS2A for which interferometer data exist. The complex organics abundance is generally higher in protostellar envelopes compared to in outflows. This suggests a chemical evolution where more CH$_3$OH ice has been converted into complex organics in the protostellar envelope compared to the more pristine ices that are released by the outflow shocks and radiation (assuming the main function of the shocks is to desorb ices rather than initiate chemistry). The SMM4 abundances are more similar to the outflow sources than envelopes around the other, much more luminous protostars, which may be due to that too small an envelope region is affected by the protostar in this case to affect the observable chemistry.

The low-mass hot cores should be further chemically evolved than the envelopes since the ices are expected to evolve during infall from the colder parts of the envelope. It is therefore unexpected that the complex abundance with respect to CH$_3$OH are smaller in the angularly resolved observations. As discussed in \S~4.1, different ice surface and ice mantle compositions may play a large role for the observed gas-phase abundance ratios. Assuming that complex molecules form mainly in the ice surface layer in cold regions, their abundances with respect to CH$_3$OH in the surface layer should increase with radiation dose e.g. between the outflows and the protostar. As long as non-thermal desorption regulates the gas-phase abundances (which should be true in most of the protostellar envelope) it is only the complex-organics enriched ice surface layer that is reflected in the gas phase abundances. Once thermal desorption becomes the dominant mechanism to release CH$_3$OH and complex organics into the gas phase, the gas phase abundances instead reflect the complex organics to CH$_3$OH ratio in the bulk of the ice mantle, which may be much lower. To confirm this scenario clearly requires a combination of laboratory experiments on surface vs bulk chemistry, astrochemical modeling and more hot core interferometric observations.

In the unresolved observations that probe both cold and warm material, HCOOCH$_3$ is generally the most abundant complex molecule, (except for CH$_3$CHO toward SMM 4) while CH$_3$OCH$_3$ or C$_2$H$_5$OH is the most abundant molecule in the low-mass hot core observations. This supports an ice chemistry scenario where complex molecules form sequentially starting with HCO-rich molecules (HCOOCH$_3$ and/or CH$_3$CHO) as long as CO ice is abundant, followed by CH$_{3/2}$-rich molecules at higher ice temperatures.

\begin{figure}[htp]
\epsscale{0.7}
\plotone{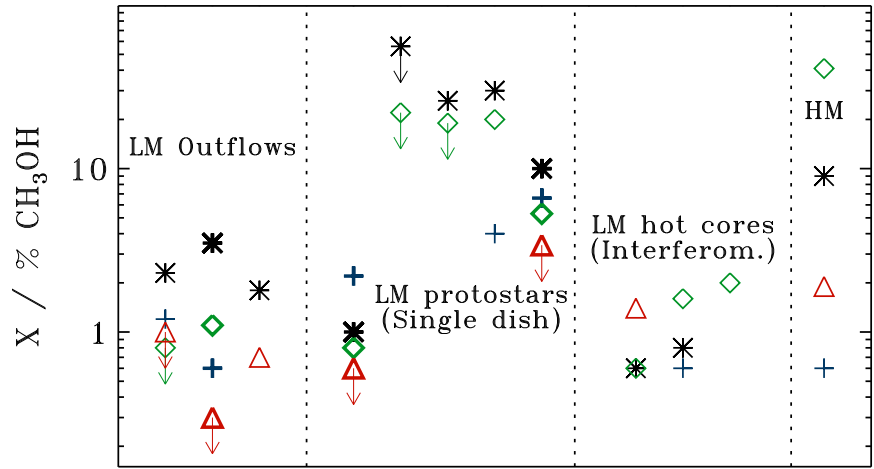}
\caption{Comparison of HCOOCH$_3$ (black stars), CH$_3$CHO (blue crosses), CH$_3$OCH$_3$ (green diamonds) and C$_2$H$_5$OH (red triangles) abundances with respect to CH$_3$OH toward low-mass outflows (B1-b, SMM4-W, L1157), protostars observed with single dish (SMM4, IRAS 4A, IRAS 4B, IRAS 16293, SMM1) and protostars observed interferometrically (IRAS~16293A, IRAS 16293B, IRAS 2A) and the average high-mass hot core abundances from \citet{Bisschop07}. \label{fig:comp_ab}}
\end{figure}

Figure \ref{fig:comp_ab} also shows how the low-mass sample of complex molecule abundances compare with average values from a single-dish survey toward high-mass hot cores \citep{Bisschop07}. Generally, the abundances are quite similar within a factor of a few toward low- and high-mass sources, but the low-mass sources do generally have lower CH$_3$OCH$_3$ abundances compared to the hot core sample. While the difference in absolute numbers may be due to the lack of applied beam dilution toward the low-mass sources, the average HCOOCH$_3$/CH$_3$OCH$_3$ ratio toward the high-mass hot cores of 0.2 is an order of magnitude lower than the ratio toward the low-mass sources observed with a single-dish telescope. The molecules in high-mass hot core have a higher temperature chemical history. The difference between the low-mass and high-mass sample further supports that the formation of complex organics is a sequential process, starting with HCO-rich organics at low temperatures. 

\section{Conclusions}

\begin{enumerate}
\item Two low-mass protostars (SMM1 and SMM4) and a low-mass outflow (SMM4-W)  in the Serpens core, known for its high CH$_3$OH ice and gas abundances,  have been probed for complex organic molecules. HCOOCH$_3$, CH$_3$CHO and CH$_3$OCH$_3$ are detected at beam-averaged abundances of 0.4--10\% with respect to CH$_3$OH.
\item The complex molecule abundances with respect to CH$_3$OH vary by an order of magnitude, with the richest complex chemistry toward the most luminous source. (HCOOCH$_3$+CH$_3$CHO)/CH$_3$OCH$_3$ abundance ratios are $\sim$3, suggestive of that cold ice chemistry dominates the complex molecule production.
\item The beam averaged abundances and excitation temperature limits indicate that the complex organics are distributed over large scales with low densities and/or low kinetic temperatures. This implies that most complex molecular emission in these sources originates from non-thermal desorption of cold ices rather than a hot core.
\item Comparison with previous observations reveal that HCOOCH$_3$ generally dominates the complex chemistry toward outflows and low-mass protostellar envelopes, consistent with experiments on CO:CH$_3$OH ice photochemistry below 25~K. In contrast, observations of   high-mass and resolved low-mass hot cores contain more CH$_3$OCH$_3$ or C$_2$H$_5$OH than HCOOCH$_3$. This supports a sequential formation of complex molecules, starting with HCO-rich molecules as long as CO ice is abundant, followed by CH$_{3/2}$-rich molecules at higher ice temperatures.
\end{enumerate}

\acknowledgments

We are grateful to the IRAM staff for their assistance and to an anonymous referee for helpful comments. Support for KIO is provided by NASA through Hubble Fellowship grant  awarded by the Space Telescope Science Institute, which is operated by the Association of Universities for Research in Astronomy, Inc., for NASA, under contract NAS 5-26555. Astrochemistry in Leiden is supported by a SPINOZA grant of the Netherlands Organization for Scientific Research (NWO). This work has benefited from research funding from the European Community's sixth Framework Programme under RadioNet R113CT 2003 5058187.

\bibliographystyle{aa}

\appendix

\section{CH$_3$OH rotational diagrams}

\begin{figure}[htp]
\epsscale{0.5}
\plotone{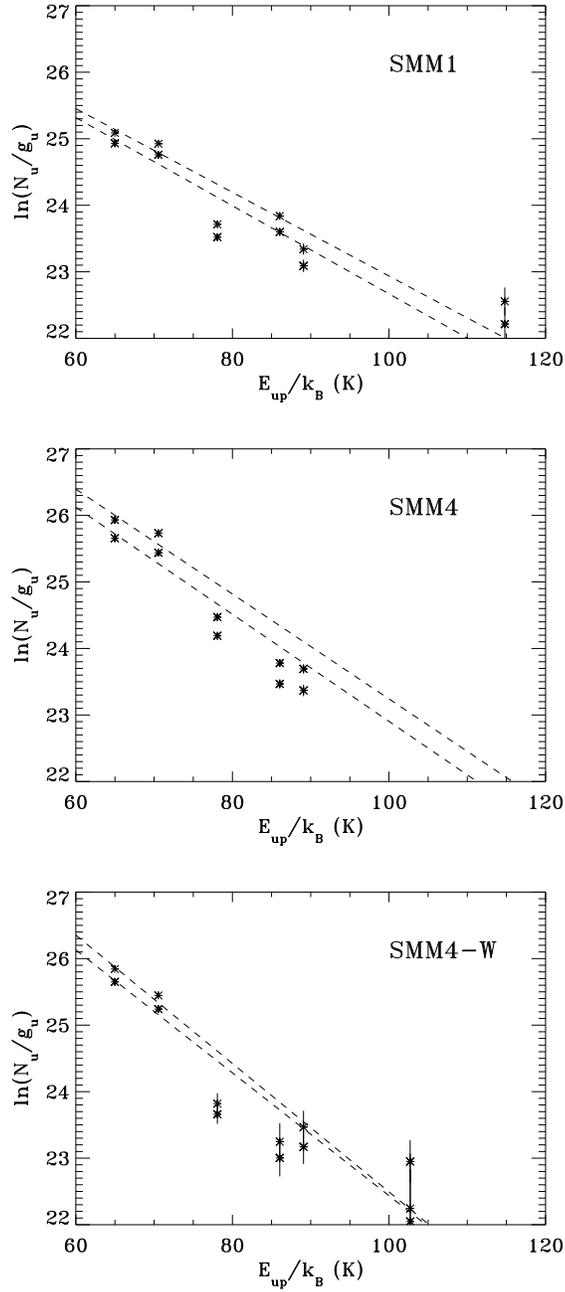}
\caption{CH$_3$OH rotational diagrams toward SMM1, SMM4 and SMM4-W for a 17" beam (upper points) and a 28" beam (lower points).\label{fig:ch3oh_smm1}}
\end{figure}

Figure \ref{fig:ch3oh_smm1} shows the rotational diagrams used to derive CH$_3$OH  rotational temperatures and beam-averaged column densities for 17" and 28" beams \citep{Goldsmith99}. The integrated CH$_3$OH intensities are from the maps presented in \citet{Kristensen10} and all other values from the JPL spectral database and CDMS. The integrated main-beam temperatures are related to the column density in the upper energy level by:

\begin{equation}
\frac{N_{\rm u}}{g_{\rm u}}=\frac{3k\int T_{\rm MB}{\rm d}V}{8\pi^3\nu\mu^2S}
\end{equation}

where $N_{\rm u}$ is the column density of the upper level, $g_{\rm u}$ is the degeneracy in the upper level, $k$ is BoltzmannÕs constant, $\nu$ is the transition frequency, $\mu$ is the dipole moment, and $S$ is the line strength. The total beam-averaged column density $N_{\rm T}$ in cm$^{-2}$ can then be computed from:

\begin{equation}
\frac{N_{\rm u}}{g_{\rm u}}=\frac{N_{\rm T}}{Q(T_{\rm rot})}e^{-E_{\rm u}/T_{\rm rot}}
\end{equation}

where $Q(T_{\rm rot})$ is the rotational partition function, and $E_{\rm u}$ is the upper level energy in K. In the rotational diagram the logarithm of $N_{\rm u}/g_{\rm u}$ is plotted versus $E_{\rm u}$ and the column density is derived from the y-axis intercept and the temperature from the slope. Figure \ref{fig:ch3oh_smm1} shows that the CH$_3$OH lines are reasonably well fit by a single temperature in all lines of sight when taking into account the line intensity uncertainties and the uncertainties in relative CH$_3$OH A and E abundances. 

There are, however, some systematic deviations from these fits, implying that the level populations cannot be characterized by single excitation temperatures perfectly. This is expected since material at different distances from the protostars and from the shock fronts are included in the beam. \citet{Kristensen10} modeled the CH$_3$OH abundances toward several Serpens cores using both the rotational diagram method, and using RATRAN and envelope models with radially symmetric temperature and density profiles. The obtained CH$_3$OH abundances were remarkably similar, within a factor of two. For the purpose of this study, where the CH$_3$OH abundances are only needed as a reference for the abundances of more complex molecules, the rotational diagram method should therefore be sufficient.

\newpage

\begin{table}
{\scriptsize
\begin{center}
\caption{Source characteristics and observational parameters\label{tab:obs}}
\bigskip

\begin{tabular}{lccc}
\hline\hline
\multicolumn{4}{c}{Source parameters}\\
\hline
Name & R.A. (J2000) & Dec. (J2000) & L$_{\rm bol}$ / L$_\odot$\\
\hline
SMM1 & 18:29:49.80&01:15:20.5&  30\\
SMM4 & 18:29:56.60& 01:13:15.1 & 5\\
SMM4-W &18:29:11.60 & 01:13:18.1 & --\\
\hline
\end{tabular}

\bigskip

\begin{tabular}{ccccc c cc}
\hline \hline
\multicolumn{8}{c}{Complex molecules observations} \\
\hline
Setting & Receiver & Frequency band (GHz) & \multicolumn{3}{c}{rms (mK)} && $\delta V$ (km s$^{-1}$) \\
\cline{4-6}
& & & SMM1 & SMM4 &SMM4-W & \\
\hline
1 	& E90 	& 88.816--88.924 	&7	& 7	&8	&&0.53\\
	&E150 	& 130.836--130.944	&10	&8	&11	&&0.35\\
2 	& E90 	& 115.465--115.573 &16	&18	&29  &&0.41\\
  	& E150 	& 146.931--147.039 &6  	&6	&10	&&0.32\\
3	&E90	&106.631--106.739	&9	&10	&\nodata	&&0.44\\
 	& E150 	& 146.817--146.923 	&9  	&11	&\nodata	&&0.32\\
\hline
\end{tabular}
\end{center}
}
\end{table}

\newpage

\begin{table}[htp]
{\scriptsize
\begin{center}
\caption{CH$_3$OH column densities and rotational temperatures for different beam sizes. \label{tab:ch3oh}}
\begin{tabular}{l cc cc}
\hline \hline
Source & \multicolumn{2}{c}{N / 10$^{14}$ cm$^{-2}$} & \multicolumn{2}{c}{Rotational T / K} \\  
&17"&28"&17"&28"\\
\hline
SMM1	&$2.5\pm0.3$&$2.5\pm0.3$&$15.9\pm0.4$&$15.1\pm0.4$\\
SMM4	&$11.5\pm1.0$&$9.4\pm1.0$&$12.7\pm0.2$&$12.4\pm0.2$\\
SMM4-W&$22\pm7$&$15\pm4$&$10.4\pm0.5$&$10.8\pm0.5$\\
\hline
\end{tabular}
\end{center}
}
\end{table}

\newpage

\begin{table*}[htp]
{\scriptsize
\begin{center}
\caption{Observed complex molecular lines with 1-$\sigma$ uncertainties in brackets and 3-$\sigma$ upper limits. \label{tab:comp}}
\begin{tabular}{lccc ccc c ccc}
\hline \hline
Molecule & Transition & Freq. & $E_{\rm low}$   & \multicolumn{3}{c}{FWHM /  km s$^{-1}$} &&  \multicolumn{3}{c}{$\int T dV$ / mK km s$^{-1}$}\\  
\cline{5-7} \cline{9-11}
& &GHz& cm$^{-1}$& SMM1 & SMM4 &SMM4-W && SMM1 & SMM4 &SMM4-W \\
\hline
HCOOCH$_3$-E & $7_{1\:6}-7_{1\:5}$ &88.8432 &9.5  &1.1[0.6]&\nodata&1.6[0.3]&&66[7]&$<$21&99[8]\\
HCOOCH$_3$-A & $7_{1\:6}-7_{1\:5}$ &88.8516 &9.5 &0.7[0.5]&\nodata&2.3[0.5]&&33[7]&$<$21&116[8]\\
\smallskip
cis-CH$_2$OHCHO & $9_{4\:6}-9_{3\:7}$	&88.8925	&21.2 & \nodata&\nodata&\nodata&&$<$21&$<$21&$<$24\\

HCOOCH$_3$-E &$14_{4\:11}-14_{3\:12}$&106.6328 &47&\nodata&\nodata&\nodata&&$<$27&$<$30&\nodata\\ 
\smallskip 
HCOOCH$_3$-A &$14_{4\:11}-14_{3\:12}$&106.6681 &47&\nodata&\nodata&\nodata&&$<$27&$<$30&\nodata\\

CH$_3$CHO-A &$6_{2\:5}-5_{2\:4}$ &115.4939 &15.9  &1.2[0.9]&1.0[0.2]&\nodata&&120[20]&120[20]&$<$78\\
C$_2$H$_5$OH &$5_{1\:5}-5_{0\:5}$ &115.4945 &48.2  &\nodata&\nodata&\nodata&&$<$50&$<$60&$<$78 \\
CH$_3$OCH$_3$ &$5_{1\:5\:2}-5_{0\:4\:2}$ &115.5440 &6.3  &\multirow{4}{*}{\nodata} &\multirow{4}{*}{\nodata} &\multirow{4}{*}{\nodata}& &\multirow{4}{*}{$<$150} &\multirow{4}{*}{$<$90} &\multirow{4}{*}{$<$156}\\
CH$_3$OCH$_3$ &$5_{1\:5\:3}-5_{0\:4\:3}$ &115.5440 &6.3  \\
CH$_3$OCH$_3$ &$5_{1\:5\:1}-5_{0\:4\:1}$ &115.5448 &6.3  \\
\smallskip
CH$_3$OCH$_3$ &$5_{1\:5\:0}-5_{0\:4\:0}$ &115.5457 &6.3  \\

C$_2$H$_5$OH &$4_{3\:2\:2}-4_{2\:3\:2}$ &130.8715 &9.3  &\nodata&\nodata&\nodata&&$<30$&$<$24&$<$27 \\
CH$_3$CHO-E &$7_{1\:7}-6_{1\:6}$ &130.8918 &14.7& \multirow{2}{*}{2.5[0.3]} &\multirow{2}{*}{2.6[0.1]} &\multirow{2}{*}{3.1[0.2]}& &\multirow{2}{*}{290[10]} &\multirow{2}{*}{560[10]} &\multirow{2}{*}{150[9]}\\
CH$_3$CHO-A &$7_{1\:7}-6_{1\:6}$ &130.8927 &14.7 \\
\smallskip
CH$_3$CH$_2$CN &$15_{0\:15}-14_{0\:14}$ & 130.9039  &	30.9 &\nodata&\nodata&\nodata&&$<30$&$<$24&$<$27 \\

CH$_3$OCH$_3$ &$5_{3\:3\:2}-5_{2\:4\:2}$ &146.8660 &13.4  & \multirow{4}{*}{\nodata} &\multirow{4}{*}{\nodata} &\multirow{4}{*}{\nodata}& &\multirow{4}{*}{$<$27} &\multirow{4}{*}{$<$33} &\multirow{4}{*}{\nodata}\\
CH$_3$OCH$_3$ &$5_{3\:3\:3}-5_{2\:4\:3}$ &146.8703 &13.4  &\\
CH$_3$OCH$_3$ &$5_{3\:3\:1}-5_{2\:4\:1}$ &146.8725 &13.4 &\\
\smallskip
CH$_3$OCH$_3$ &$5_{3\:3\:0}-5_{2\:4\:0}$ &146.8773 &13.4 & \\

HCOOCH$_3$-E &$12_{3\:10}-11_{3\:9}$ &146.9777 &31.3  &\nodata&\nodata&\nodata&&$<$63&$<$490&$<$140 \\
HCOOCH$_3$-A &$12_{3\:10}-11_{3\:9}$ &146.9880 &31.2  &\nodata&\nodata&\nodata&&$<$18&$<$35&$<$30  \\
CH$_3$OCH$_3$ &$7_{1\:7\:2}-6_{0\:6\:2}$ &147.0242 &13.2   & \multirow{4}{*}{2.4} &\multirow{4}{*}{2.2} &\multirow{4}{*}{4.4}& &\multirow{4}{*}{84[6]} &\multirow{4}{*}{52[6]} &\multirow{4}{*}{120[10]}\\
CH$_3$OCH$_3$ &$7_{1\:7\:3}-6_{0\:6\:3}$ &147.0242 &13.2   \\
CH$_3$OCH$_3$ &$7_{1\:7\:1}-6_{0\:6\:1}$ &147.0249 &13.2  \\
CH$_3$OCH$_3$ &$7_{1\:7\:0}-6_{0\:6\:0}$ &147.0256 &13.2 \\
\hline
\end{tabular}
\end{center}
}
\end{table*}

\newpage

\begin{table*}[htp]
{\scriptsize
\begin{center}
\caption{Derived column densities and abundances using CH$_3$OH rotational temperatures \label{tab:cols}}

\begin{tabular}{lccc ccc c ccc}
\hline \hline
Molecule & Transition & beam size   & \multicolumn{3}{c}{N /  10$^{13}$ cm$^{-2}$} &&  \multicolumn{3}{c}{\% CH$_3$OH}\\  
\cline{4-6} \cline{8-10}
& &"& SMM1 & SMM4 &SMM4-W && SMM1 & SMM4 &SMM4-W \\
\hline

HCOOCH$_3$-E & $7_{1\:6}-7_{1\:5}$ & 28&1.7&$<$0.5&2.4 &&6.7&$<$0.5&1.6\\
HCOOCH$_3$-A & $7_{1\:6}-7_{1\:5}$ & 28&0.8&$<$0.5&2.9 &&3.3&$<$0.5&1.9\\
\smallskip
HCOOCH$_3$-A &$12_{3\:10}-11_{3\:9}$ &  17&$<$1.4&$<$4.6&$<$7.2 &&$<$5.7&$<$4.0&$<$3.3\\

CH$_3$CHO-A &$6_{2\:5}-5_{2\:4}$ &21&0.9&1.0&1.0&&3.7&0.9&0.4\\
CH$_3$CHO-E &$7_{1\:7}-6_{1\:6}$ &21& 0.7&1.5&0.5 &&2.9&1.3&0.2\\%
\smallskip
CH$_3$CHO-A &$7_{1\:7}-6_{1\:6}$ & 19& 0.7&1.5&0.5 &&2.9&1.3&0.2\\

CH$_3$OCH$_3$ &$5_{1\:5\:2}-5_{0\:4\:2}$ &\multirow{4}{*}{21}  &\multirow{4}{*}{$<$2.1} &\multirow{4}{*}{$<$1.2} &\multirow{4}{*}{$<$2.0}& &\multirow{4}{*}{$<$8.4} &\multirow{4}{*}{$<$1.0} &\multirow{4}{*}{$<$0.9}\\
CH$_3$OCH$_3$ &$5_{1\:5\:3}-5_{0\:4\:3}$ & \\
CH$_3$OCH$_3$ &$5_{1\:5\:1}-5_{0\:4\:1}$ & \\
CH$_3$OCH$_3$ &$5_{1\:5\:0}-5_{0\:4\:0}$ & \\

CH$_3$OCH$_3$ &$5_{3\:3\:2}-5_{2\:4\:2}$ &\multirow{4}{*}{17}  &\multirow{4}{*}{$<$0.9} &\multirow{4}{*}{$<$1.2} &\multirow{4}{*}{\nodata}& &\multirow{4}{*}{$<$3.7} &\multirow{4}{*}{$<$1.1} &\multirow{4}{*}{\nodata}\\
CH$_3$OCH$_3$ &$5_{3\:3\:3}-5_{2\:4\:3}$ &\\
CH$_3$OCH$_3$ &$5_{3\:3\:1}-5_{2\:4\:1}$ &\\
CH$_3$OCH$_3$ &$5_{3\:3\:0}-5_{2\:4\:0}$ & \\

CH$_3$OCH$_3$ &$7_{1\:7\:2}-6_{0\:6\:2}$ &\multirow{4}{*}{17} &\multirow{4}{*}{1.3} &\multirow{4}{*}{0.9} &\multirow{4}{*}{2.4}& &\multirow{4}{*}{5.3} &\multirow{4}{*}{0.8} &\multirow{4}{*}{1.1}\\
CH$_3$OCH$_3$ &$7_{1\:7\:3}-6_{0\:6\:3}$ &  \\
CH$_3$OCH$_3$ &$7_{1\:7\:1}-6_{0\:6\:1}$ & \\
\smallskip
CH$_3$OCH$_3$ &$7_{1\:7\:0}-6_{0\:6\:0}$ & \\

C$_2$H$_5$OH &$4_{3\:2\:2}-4_{2\:3\:2}$ &19&$<$0.9&$<$0.6&$<$0.8&&$<$3.4&$<$0.6&$<$0.3 \\

\hline
\end{tabular}
\end{center}
}
\end{table*}

\newpage

\begin{table*}[ht]
{\scriptsize
\begin{center}
\caption{CH$_3$OH data and beam averaged complex molecule abundances toward low-mass sources\label{tab:lowmass}}

\begin{tabular}{l ccc c cccc c}
\hline \hline
Source & \multicolumn{3}{c}{CH$_3$OH} &&HCOOCH$_3$&CH$_3$CHO&CH$_3$OCH$_3$&C$_2$H$_5$OH &Refs\tablenotemark{a}\\  
\cline{2-4} 
& N / 10$^{14}$ cm$^{-2}$ & beam / " & T$_{\rm rot}$ / K&&  \multicolumn{4}{c}{\% CH$_3$OH}\\
\hline
\smallskip
B1-b core &4.7&19&10[5]						&&2.3&1.2&$<$0.8&$<$1.0 &1\\

SMM4-W	&22[7]&22.5&11[1]					&&3.5&0.6&1.1&$<$0.4\\
\smallskip
L1157 outflow &15&10.5&12[2]				&&1.8&\nodata&\nodata&0.7 &2,3\\

SMM1	&2.5[0.3]&22.5&16[1]				&&10.0&6.6&5.3&$<$3.4\\
SMM4	&10.5[1.0]&22.5&13[1]				&&$<$1.0&2.2&0.8&$<$0.6\\
NGC1333 IRAS 4A\tablenotemark{b} &5.1&10&24	&&56 & \nodata &$<$22 &\nodata&4,5,6\\
NGC1333 IRAS 4B\tablenotemark{b} &3.5&10&34	&&26 & \nodata &$<$19 &\nodata&4,5,6\\
\smallskip
IRAS 16293 env.\tablenotemark{b}&8.8&20&85&&30 &4&20 &\nodata&6,7,8\\

IRAS 16293 A\tablenotemark{c}   	&1.1$\times10^4$&1.3''$\times$2.7'' &100	&&0.6&$<$0.02&0.6&1.4&9,10\\
IRAS 16293 B\tablenotemark{c}  	&5.0$\times10^3$&1.3''$\times$2.7'' &100	&&0.8&0.6&1.6&\nodata&9,10\\
NGC1333 IRAS 2A\tablenotemark{c} &2$\times10^4$&1.8''$\times$1.0''&$>$75~K	&&\nodata&\nodata&2&\nodata&11,12\\
\hline
\end{tabular}
\end{center}
}
\tablenotetext{a}{1. \citet{Oberg10a}, 2. \citet{Bachiller97}, 3. \citet{Arce08}, 4. \citet{Maret05}, 5. \citet{Bottinelli07}, 6.\citet{Herbst09}, 7. \citet{vanDishoeck95}, 8. \citet{Cazaux03}, 9. \citet{Kuan04}, 10. \citet{Bisschop08}, 11. \citet{Huang05}, 12. \citet{Jorgensen05}.}
\tablenotetext{b}{The complex abundances values were calculated assuming the {\it same} beam dilution for CH$_3$OH and the complex species, which should correspond to comparing beam-averaged abundances since the beams are of similar size \citep{Herbst09}. }
\tablenotetext{c}{The complex abundances are also from interferometric observations. }
\end{table*}

\newpage

\end{document}